# Percolation Modeling of Conductance of Self-Healing Composites

**Alexander Dementsov** and **Vladimir Privman**\*

Center for Advanced Materials Processing, and
Department of Physics, Clarkson University,
Potsdam, New York 13699, USA

### Abstract

We explore the conductance of self-healing materials as a measure of the material integrity in the regime of the onset of the initial fatigue. Continuum effective-field modeling and lattice numerical simulations are reported. Our results illustrate the general features of the self-healing process: The onset of the material fatigue is delayed, by developing a plateau-like time-dependence of the material quality. We demonstrate that in this low-damage regime, the changes in the conductance and similar transport/response properties of the material can be used as measures of the material quality degradation.

---

\* www.clarkson.edu/Privman

Recently a significant research effort has been devoted to the design of "smart materials." In particular, self-healing composites [1-10] can restore their mechanical properties with time or at least reduce material fatigue caused by the formation of microcracks. It is expected that microcracks propagating through such materials can break embedded capsules/fibers which contain the healing agent — a "glue" that heals/delays further microcrack development — thus triggering the self-healing mechanism. In recent experiments [1,7-10], an epoxy (polymer) was studied, with embedded microcapsules containing a healing agent. Application of a periodic load on a specimen with a crack, induced rupture of microcapsules [1]. The healing glue was released from the damaged microcapsules, permeated the crack, and a catalyst triggered a chemical reaction which re-polymerized the crack.

Defects of nanosizes are randomly distributed throughout the material. Mechanical loads during the use of the material then cause formation of craze fibrils along which microcracks develop. This leads to material fatigue and, ultimately, degradation. Triggering self-healing mechanism at the *nanoscale* might offer several advantages [10] for a more effective prevention of growth of microcracks. Indeed, it is hoped [10] that nanoporous fibers with glue will heal smaller damage features, thus delaying the material fatigue at an earlier stage than larger capsules [1,9] which basically re-glue large cracks. Furthermore, on the nanoscale, the glue should be distributed/mixed with the catalyst more efficiently because transport by diffusion alone will be effective [10,11], thereby also eliminating the need for external UV irradiation [9], etc.

Theoretical and numerical modeling of self-healing materials are only in the initiation stages [10,12,13]. Many theoretical works and numerical simulations [14-17] consider formation and propagation of large cracks which, once developed, can hardly be healed by an embedded nano-featured capsules. Therefore, we have proposed [10] to focus the modeling program on the time dependence of a gradual formation of damage (fatigue) and its manifestation in material composition, as well as its healing by nanoporous fiber rupture and release of glue.

We will shortly formulate rate equations [10] for such a process. In addition to continuum rate equations for the material composition, numerical modeling can yield useful information on the structure, and, later in this article, we report results of Monte Carlo simulations. We also point out that the calculated material composition and structure must be related to macroscopic properties that are experimentally probed. The relation between composite materials composition and properties is an important and rather broad field of research [18].

Recently, it has been demonstrated experimentally [19] that a rather dilute network of carbon nanotubes, incorporated in the epoxy matrix, can provide a percolation cluster the conductance of which can not only reflect the degree of the fatigue of the material but also shows promise for probing the self-healing process. The main purpose of the present article is to initiate continuum effective-field, as well as numerical lattice modeling of percolation properties for materials with self-healing.

Different transport properties can be used to probe material integrity (damage accumulation due to the formation of cracks). These include thermal conductivity [20,21], photoacoustic waves [22,23], electrical conductivity [19,24-27]. Generally, transport properties can be highly nonlinear as functions of the degree of damage. For example, the conductance can sharply drop to zero if the conducting network density drops below the percolation threshold. However, for probing the initial fatigue, in the regime of low levels of damage, one expects most transport properties to decrease proportionately to the damage.

Let us summarize our recently proposed model [10] of the material composition in the continuum rate equation approach. We denote by $u(t)$ the fraction of material that is undamaged, by $g(t)$ the fraction of material consisting of glue-carrying capsules, by $d(t)$ the fraction of material that is damaged, and by $b(t)$ the fraction of material with broken capsules, so that we have

$$u(t) + g(t) + d(t) + b(t) = 1. \tag{1}$$

We consider the regime of small degree of degradation of the material, i.e., we assume that at least for small times, $t$, we have $u(t) \approx 1$, whereas $d(t)$, $b(t)$ and $g(t)$ are relatively small. In fact, $b(0) = 0$.

For the purposes of simple modeling, we assume that *on average* the capsules degrade with the rate $P$, which is somewhat faster than the rate of degradation of the material itself due to its continuing use (fatigue), $p$, i.e., $P > p$. The latter assumption was made to mimic the expected property that a significant amount of microcapsules embedded in the material may actually weaken its mechanical properties and, were it not for their healing effect, reduce its usable lifetime (though it was noted [1,11] that a small amount of microcapsules actually increased the epoxy toughness); the density of the "healing" microcapsules is one of the important system parameters to optimize in any modelling approach. Thus, we approximately take

$$\dot{g}(t) = -Pg(t), \qquad \text{yielding} \qquad g(t) = g(0)e^{-Pt}. \tag{2}$$

One can write a more complicated rate equation for $g(t)$, but the added, nonlinear terms are small in the considered regime.

However, for the fraction of the undamaged material, we cannot ignore the second, nonlinear term in the relation

$$\dot{u}(t) = -pu(t) + H(t). \tag{3}$$

Here we introduced the healing efficiency, $H(t)$, which can be approximated by the expression

$$H(t) \propto d(t)g(t) \times (\text{volume healed by one capsule}). \tag{4}$$

The healing efficiency is proportional to the fraction of glue capsules, as well as to the fraction of the damaged material, because that is where the healing process is effective. The latter will be approximated by $d(t) \approx 1 - u(t)$, which allows us to obtain a closed equation for $u(t)$. Indeed, in Eq. (3) we can now use

$$H(t) = Ae^{-Pt}[1 - u(t)]. \tag{5}$$

The healing efficiency is controlled by the parameter

$$A \propto g(0) \times (\text{volume healed by one capsule}). \tag{6}$$

While the model just formulated is quite simple, "minimal," and many improvements can be suggested, it has the advantage of offering an exact solution,

$$u(t) = u(0)e^{-pt - AP^{-1}(1-e^{-Pt})} + Ae^{-pt + AP^{-1}e^{-Pt}} \int_0^t d\tau \, e^{-(P-p)\tau - AP^{-1}e^{-P\tau}}. \tag{7}$$

This result is illustrated by the solid curves in Fig. 1, where we set $u(0) = 1$ for simplicity. The main feature observed is that even when the healing efficiency parameter $A$ is rather small (here 0.02) but nonzero, the decay of the fraction of the undamaged material is delayed for some interval of time. This represents the self-healing effect persisting until the glue capsules are used up.

Equation (6) suggests that an important challenge in the design of self-healing materials will be to have the healing effect of most capsules cover volumes much larger than a capsule, in order to compensate for a relatively small value of $g(0)$, which is the fraction of the material volume initially occupied by the glue-filled capsules. Since the glue cannot "decompress," its healing action, after it spreads out and solidifies, should have a relatively long-range stress-relieving effect in order to prevent further crack growth over a large volume.

The present simple continuum modeling cannot address the details of the morphological material properties and glue transport; numerical simulations will be needed to explore this issue. It is interesting to note that most *material properties* will also depend on the specific morphological assumptions; their derivation within various approximation schemes, will require more information than that provided by the average, "effective field" approximate "materials quality" measures such as $u(t)$. Here we are interested, specifically, in the material conductivity.

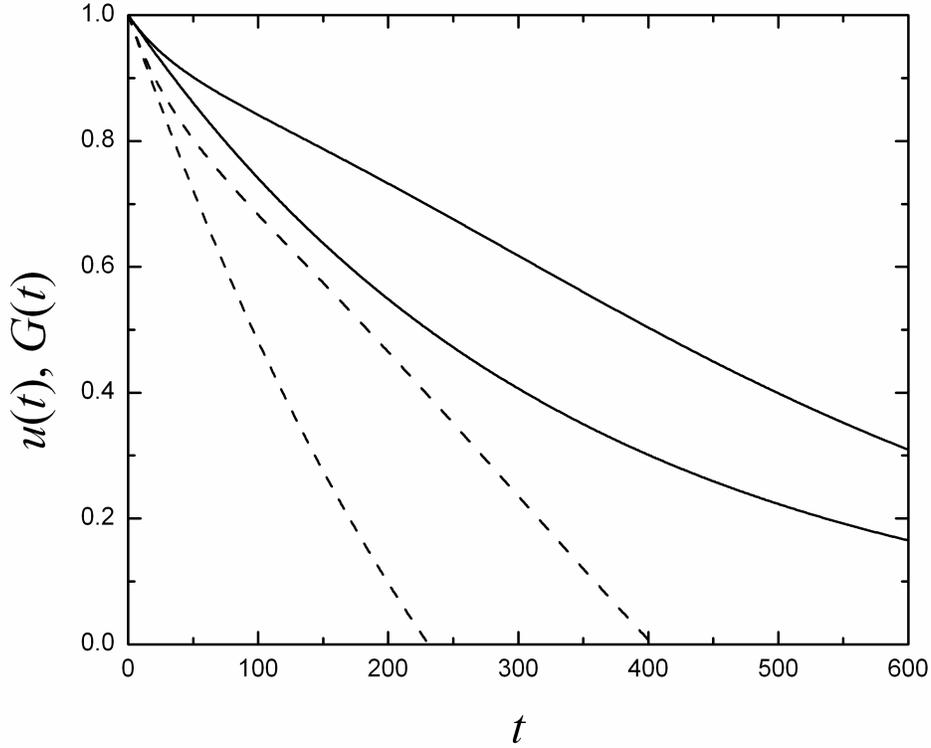

**Figure 1:** The solid curves illustrate the fraction of the undamaged material, $u(t)$, calculated according to Eq. (7) with $A = 0.02$, $p = 0.003$, $P = 0.008$ — the top curve, and without self-healing: $A = 0$, $p = 0.003$ — the bottom curve. The dashed curves illustrate the behavior of the mean-field conductance for these two cases, respectively, with the conductance decreasing slower with self-healing present, eventually reaching zero at the percolation transition at $u = 0.5$, see Eq. (8).

---

Since our numerical calculations reported below, assume square lattice (coordination number $z = 4$) bond percolation, the conductance, $G(t)$, shown as the dashed curves in Fig. 1, was calculated by using the bond-percolation mean-field formula [28],

$$G(t) = \max\left[\,1 - z\frac{1-u(t)}{z-2},\,0\,\right] = \max\left[\,2u(t) - 1,\,0\,\right]. \qquad (8)$$

Here the conductance is normalized to have $G(t=0) = 1$, and for simplicity we assumed that the bond percolation probability is given by $u(t)$, i.e., we consider the situation when the conductance of the healthy/healed material is maximal, whereas the other areas do not conduct at all. In the regime of a relatively low damage, which is likely the only one of practical interest, and also the one where the mean-field expressions are accurate, we note that the conductance provides a convenient, proportional measure of the material degradation,

$$G(0) - G(t) \simeq K[u(0) - u(t)], \qquad (9)$$

where the constant $K = z/(z-2)$ depends on the microscopic details of the material conductivity. Here $K = 2$, but in practical situations this parameter can be fitted from experimental data.

In order to further explore the self-healing process, we carried out Monte Carlo simulations on square lattices of varying sizes, with periodic boundary conditions. All the bonds in the lattice were initially present, and the healing cells were a small fraction of the lattice (square) unit cells, distributed uniformly over the lattice, with the built in constraint that they do not touch each other (including no corner contact). Our simulations reported here, were carried out with this fraction $g(0) = 0.15$, i.e., the probability that a cell was designated glue-carrying was 15%.

At times $t > 0$, bonds were randomly broken with the probability (rate per unit time) $p$ for ordinary bonds, and $P\,(>p)$ for bonds of healing cells (those with "glue"). If at least two bonds are broken in a healing cell, the glue leaks out and restores broken bonds. Here we assumed *local* healing, with the glue only spreading to the 8 neighboring cells before solidifying, thus restoring all the 24 bonds of the $3 \times 3$ square group of cells that includes the healing cell as its center. Furthermore, once the glue leaks out, the healing cell becomes inactive, but its bonds still have the larger probability, $P$, to be re-broken.

We further assumed that all the original or healed bonds have the same, maximal conductance, whereas all the broken bonds do not conduct at all. Since the periodic boundary conditions induce a torus geometry, the conductance of a system of $N \times N$ square cells, was calculated between two parallel lines, each $N$ lattice bonds long (which were really circles due to periodicity), at the distance $N/2$ from each other, by using a standard algorithm [29]. Note that these two lines are connected by two equal-size system halves (we took $N$ even for simplicity), and the conductivities via these two pathways were included in the overall calculation. Our typical results are illustrated in Fig. 2, where we plot the number (fraction) of unbroken bonds, $n(t)$, with initially $n(0) = 1$, as well as the normalized conductance, $G(t)$.

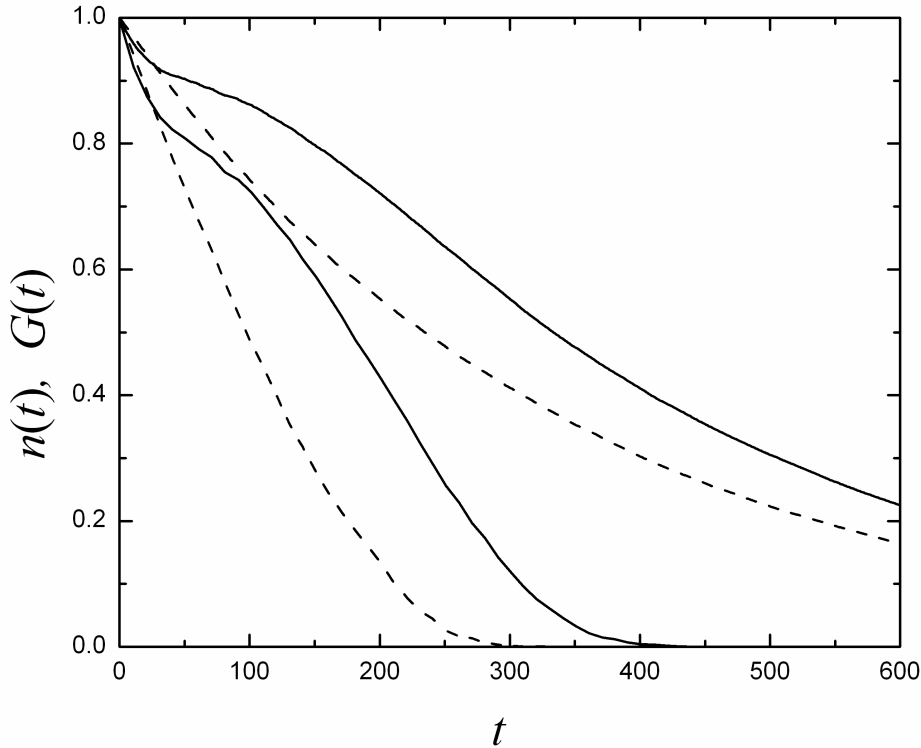

**Figure 2:** The solid curves illustrate the fraction of the unbroken bonds, $n$ (top curve), and the normalized conductance, $G$ (bottom curve), for the following choice of the parameters: $N = 32$, $p = 0.003$, $P = 0.008$, and the initial fraction of the healing cells 15%. The dashed curves illustrate similar results with the same parameters but with no healing cells. The data were averaged over 40 Monte Carlo runs for the case with self-healing, and over 20 runs for the case without self-healing.

---

The two lower curves in Fig. 2, showing the conductance with and without self-healing, do not vanish at finite times due to finite-size effects [30]. In fact, without self-healing the simulation of the conductance is just the ordinary numerical evaluation for square-lattice uncorrelated bond percolation. As the lattice size increases, the finite-lattice conductance, as well as other percolation properties, develop critical-point behavior at the percolation transition that occurs, for this particular morphology, when the fraction of the broken bonds reaches 0.5. Thus, the conductance, $G(t)$, shows significant lattice-size dependence even without self-healing, as illustrated in Fig. 3, whereas the fraction of the unbroken bonds, $n(t) = e^{-pt}$, not shown in the figure, has no size dependence.

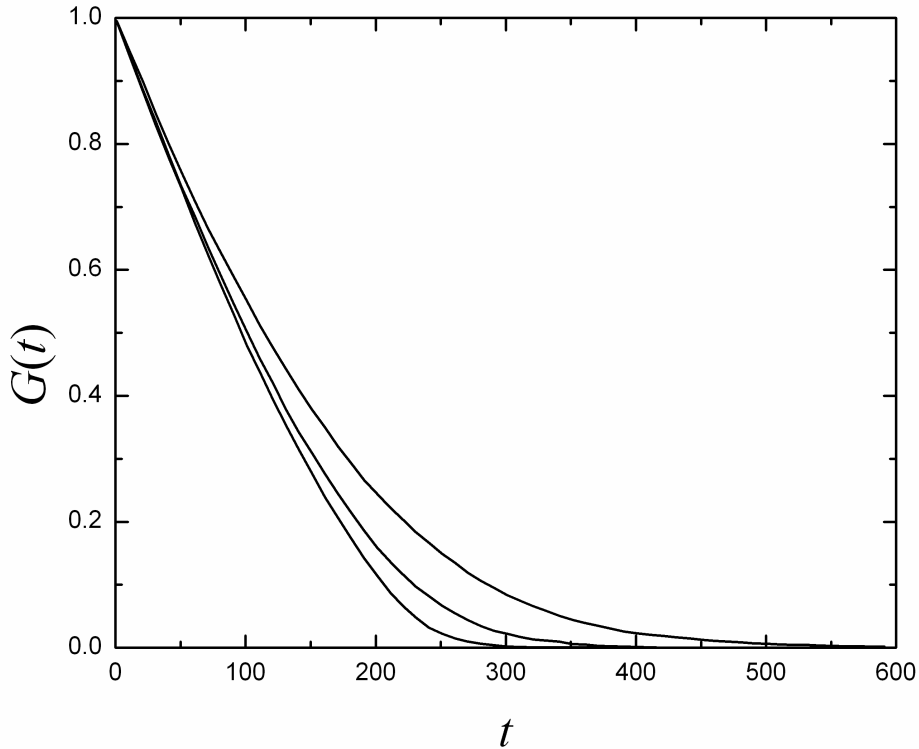

**Figure 3:** Size dependence of the conductance without self-healing, with otherwise the same parameters as in Fig. 2. From top to bottom, the results shown correspond to lattice sizes $N = 8$, 16, 32. The data were averaged over 500, 100 and 20 Monte Carlo runs, respectively. Note that the percolation transition occurs at $t = (\ln 2)/p \simeq 231$, from which time value on, the $N \to \infty$ limiting value of the conductance is zero.

---

Results with self-healing, for the size dependence of the conductance, are shown in Fig. 4. We point out that the fraction of the unbroken bonds also has some variation with $N$ in this case. However, the differences in $n$-values are too small to be displayed in the figure. (The size dependence of $n(t;N)$ might become quite pronounced and interesting when the self-healing process is non-local, as discussed in [10].) For most practical purposes the self-healing process will be of interest as long as the material fatigue is small, i.e., in the regime of the initial plateau that develops in properties such as $n(t)$, or $u(t)$ in the continuum model. Therefore, we did not attempt to study in detail the percolation transition for the conductance, which in this case should be a variant of some sort of a correlated bond percolation, though the universality class is likely not changed.

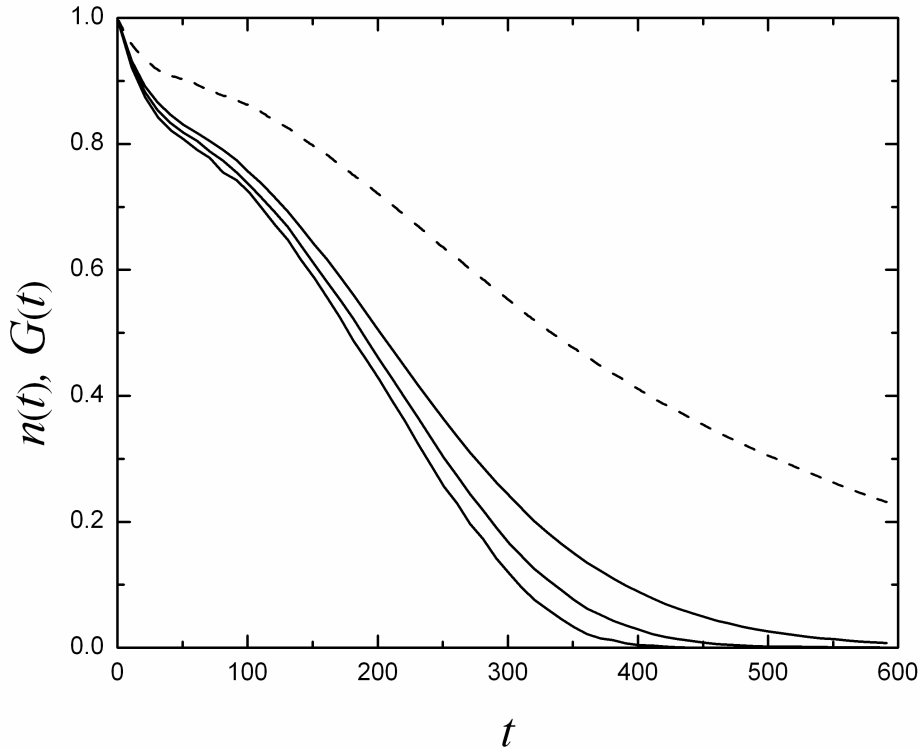

**Figure 4:** Size dependence of the conductance (the solid curves) with self-healing, with the same parameters as in Figs. 2 and 3. From top to bottom, the solid curves correspond to lattice sizes $N = 8$, 16, 32. The data were averaged over 2000, 400 and 40 Monte Carlo runs, respectively. The dashed curve shows the fraction of the healthy bonds, the size-dependence of which leads to variations too small to be shown on the scale of the vertical axis in this plot (the curve shown is for $N = 32$).

Let us now discuss the extent to which the continuum model can fit the results for the lattice model. Now, without self-healing, the mean-field approximation can provide rather accurate results for the conductance, except perhaps right near the percolation transition [31], as illustrated in Fig. 5. With self-healing, the situation is less consistent. The numerical lattice-model result (for our largest $N = 32$), is compared to the continuum model expression with varying $A$, in Fig. 6.

While, especially for larger values of $A$, the continuum model curves show all the features of the self-healing conductance, including the initial drop followed by "shoulder," the overall agreement is at best only qualitative. Thus, using $A$ as the adjustable parameter, one cannot achieve a quantitatively accurate fit of the lattice-model data. We note that the continuum model considered, should be viewed as "minimal" in that it represents the simplest possible set of assumptions that yield the self-healing behavior and also offer exact solvability. Specifically, the continuum model assumes that the initial fraction of the glue-carrying capsules is very small, and the finite healing efficiency is achieved by each cell healing a large volume, see the discussion in connection with Eq. (6). On the other hand, to have a full-featured self-healing behavior, in the lattice case with short-range healing, we had to take the initial fraction of the healing cells at least of order 10% (we took 15% in our simulations). Thus, for better-quality fit the continuum model will have to be modified, and will be more complicated, involving more than one quantity (now we only consider $u(t)$, for which we obtain a closed equation) and likely nonlinear equations. We plan to consider this in our future work.

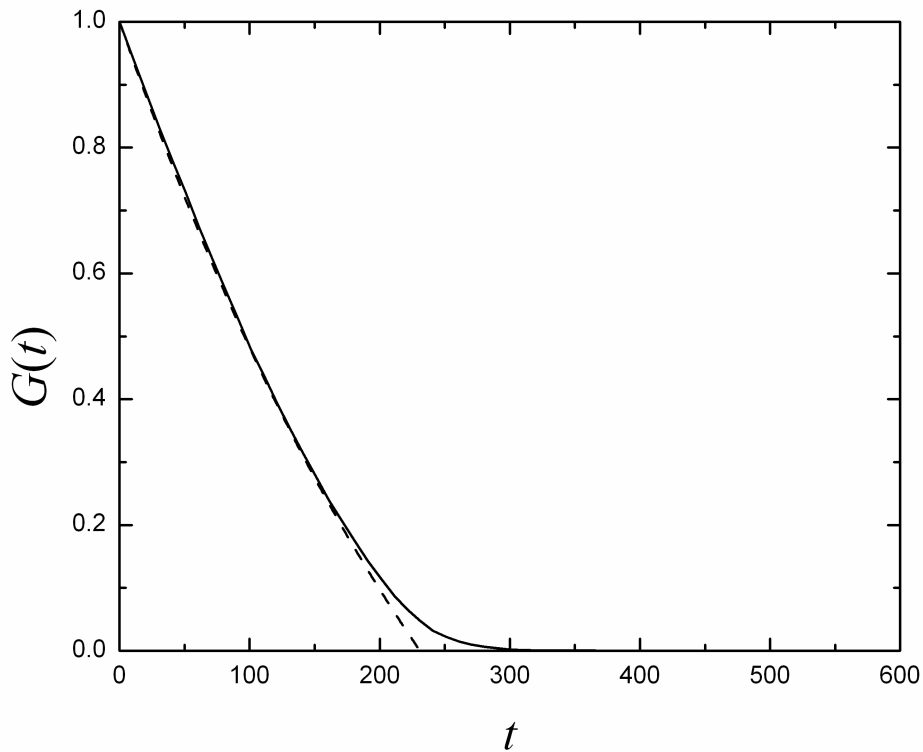

**Figure 5:** The dashed curve shows the mean-field approximation for the conductance without self-healing, calculated according to Eq. (8), with $p = 0.003$. The solid curve is the $N = 32$ lattice-model result, as in Fig. 3.

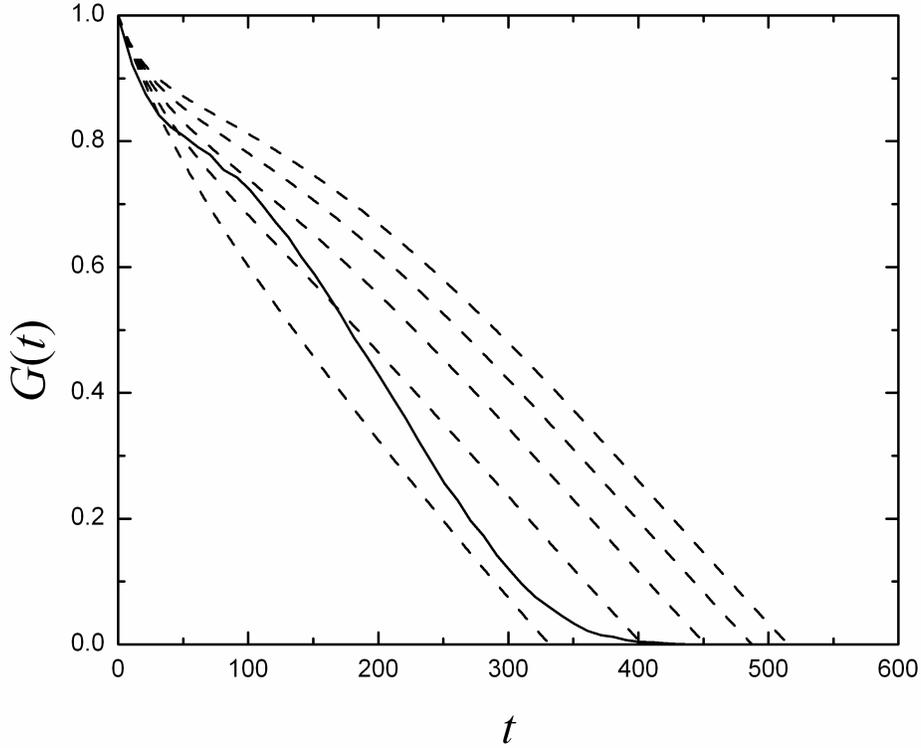

**Figure 6:** The dashed curves show the conductance calculated according to the continuum model, for $p = 0.003$ and $P = 0.008$, with, from bottom to top, $A = 0.01$, 0.02, 0.03, 0.04, 0.05. The solid curve is the $N = 32$ lattice-model result with self-healing, as in Fig. 4.

In summary, we explored the conductance of self-healing materials, with several assumptions that include short-range healing, conductivity being directly proportional to the local material "health," and the use of simple effective-field continuum model, as well as two-dimensional square lattice numerical simulations. While our assumptions may have to be modified for different, more realistic situations, our results illustrate the general features of the self-healing process. Specifically, the onset of the material fatigue is delayed, by developing a plateau-like time-dependence of the material quality at initial times. In this regime, the changes in the conductance, and likely in most other transport/response properties of the material that can be experimentally probed, measure the material quality degradation proportionately, whereas for larger damage at later times, transport properties may undergo dramatic changes, such as the vanishing of the conductance in our case, and they might not be good measures of the material integrity.

We wish to thank Dr. D. Robb for bringing reference [19] to our attention and for helpful discussions, and we acknowledge support of this research by the US-ARO under grant W911NF-05-1-0339 and by the US-NSF under grant DMR-0509104.

**References**

1. S. R. White, N. R. Sottos, P. H. Geubelle, J. S. Moore, M. R. Kessler, S. R. Sriram, E. N. Brown and S. Viswanathan, Nature **409**, 794 (2001).

2. C. Dry, Composite Structures **35**, 263 (1996).

3. B. Lawn, *Fracture of Brittle Solids* (Cambridge University Press, Cambridge, 1993), Chapter 7.

4. C. M. Dry and N. R. Sottos, Proc. SPIE **1916**, 438 (1996).

5. E. N. Brown, N. R. Sottos and S. R. White, Exper. Mech. **42**, 372 (2002).

6. Y. Kievsky and I. Sokolov, IEEE Trans. Nanotech. **4**, 490 (2005).

7. E. N. Brown, S. R. White and N. R. Sottos, J. Mater. Sci. **39**, 1703 (2004).

8. M. Zako and N. Takano, J. Intel. Mater. Syst. Struct. **10**, 836 (1999).

9. J. W. C. Pang and I. P. Bond, Compos. Sci. Tech. **65**, 1791 (2005).

10. V. Privman, A. Dementsov and I. Sokolov, J. Comput. Theor. Nanosci. **4**, 190 (2007).

11. I. Sokolov, private communication.

12. S. R. White, P. H. Geubelle and N. R. Sottos, *Multiscale Modeling and Experiments for Design of Self-Healing Structural Composite Materials*, US Air Force research report AFRL-SR-AR-TR-06-0055 (2006).

13. J. Y. Lee, G. A. Buxton and A. C. Balazs, J. Chem. Phys. **121**, 5531 (2004).

14. S. Hao, W. K. Liu, P. A. Klein and A. J. Rosakis, Int. J. .Solids Struct. **41**, 1773 (2004).

15. H. J. Herrmann, A. Hansen and S. Roux, Phys. Rev. **B 39**, 637 (1989).

16. M. Sahimi and S. Arbabi, Phys. Rev. **B 47**, 713 (1993).

17. J. Rottler, S. Barsky and M. O. Robbins, Phys. Rev. Lett. **89**, 148304 (2002).

18. G. W. Milton, *The Theory of Composites* (Cambridge University Press, Cambridge, 2001), Chapter 10.

19. E. T. Thostenson and T.-W. Chou, Adv. Mater. **18**, 2837 (2006).

20. I. Sevostianov, Int. J. Eng. Sci. **44**, 513 (2006).

21. I. Sevostianov and M. Kachanov, *Connections Between Elastic and Conductive Properties of Heterogeneous Materials*, preprint (2007).

22. M. Navarrete, M. Villagrán-Munizb, L. Poncec and T. Flores, Opt. Lasers Eng. **40**, 5 (2003).

23. A. S. Chekanov, M. H. Hong, T. S. Low and Y. F. Lu, IEEE Trans. Magn. **33**, 2863 (1997).

24. K. Schulte and C. Baron, Compos. Sci. Tech. **36**, 63 (1989).

25. I. Weber and P. Schwartz, Compos. Sci. Tech. **61**, 849 (2001).

26. M. Kupke, K. Schulte and R. Schüler, Compos. Sci. Tech. **61**, 837 (2001).

27. R. Schueler, S. P. Joshi and K. Schulte, Compos. Sci. Tech. **61**, 921 (2001).

28. S. Kirkpatrick, Rev. Mod. Phys. **45**, 574 (1973).

29. H. A. Knudsen and S. Fazekas, J. Comput. Phys. **211**, 700 (2006).

30. V. Privman, *Finite Size Scaling and Numerical Simulation of Statistical Systems* (World Scientific, Singapore, 1990).

31. H. E. Stanley, *Introduction to Phase Transitions and Critical Phenomena* (Oxford University Press, Oxford, 1993).